\documentclass{PoS}

\title{Finite-volume effects in the hadronic vacuum polarization}

\ShortTitle{Finite-volume effects in the hadronic vacuum polarization}

\author{Christopher Aubin\\
        Dept. of Physics, Fordham Univ., Bronx, New York, NY 10458, USA\\
        E-mail: \email{caubin@fordham.edu}}

\author{Thomas Blum\\
        Dept. of Physics, Univ. of Connecticut, Storrs, CT 06269, USA\\
        E-mail: \email{tblum@phys.uconn.edu}}

\author{Peter Chau\\
        Dept. of Physics and Astronomy, San Francisco State Univ., San Francisco, CA 94132, USA\\
        \phantom{E-mail{...}}}

\author{\speaker{Maarten Golterman}\\
        Dept. of Physics and Astronomy, San Francisco State Univ., San Francisco, CA 94132, USA\\
        E-mail: \email{maarten@sfsu.edu}}

\author{Santiago Peris\\
        Dept. of Physics, Univ. Aut\`onoma de Barcelona, E-08193 Bellaterra, Barcelona, Spain\\
        E-mail: \email{peris@ifae.es}}
        
\author{Cheng Tu\\
        Dept. of Physics, Univ. of Connecticut, Storrs, CT 06269, USA\\
        E-mail: \email{cheng.tu@uconn.edu}}

\abstract{We investigate finite-volume effects in the hadronic vacuum polarization, with an eye toward the corresponding systematic error in the muon anomalous magnetic moment. While it is well known that leading-order chiral perturbation theory does not provide a good description of the hadronic vacuum polarization, it turns out that it gives a much better representation of finite-volume effects. Indications are that finite-volume effects cannot be ignored when the aim is a few percent level accuracy for the hadronic contribution to the muon anomalous magnetic moment, even when $m_\pi L \sim 4$ and $m_\pi \sim 200$ MeV.}

\FullConference{The 33rd International Symposium on Lattice Field Theory\\
		14 -18 July 2015\\
		Kobe International Conference Center, Kobe, Japan*}

\begin{document}

\section{Introduction}
\begin{figure}[t]
\centering
\includegraphics[width=2.9in]{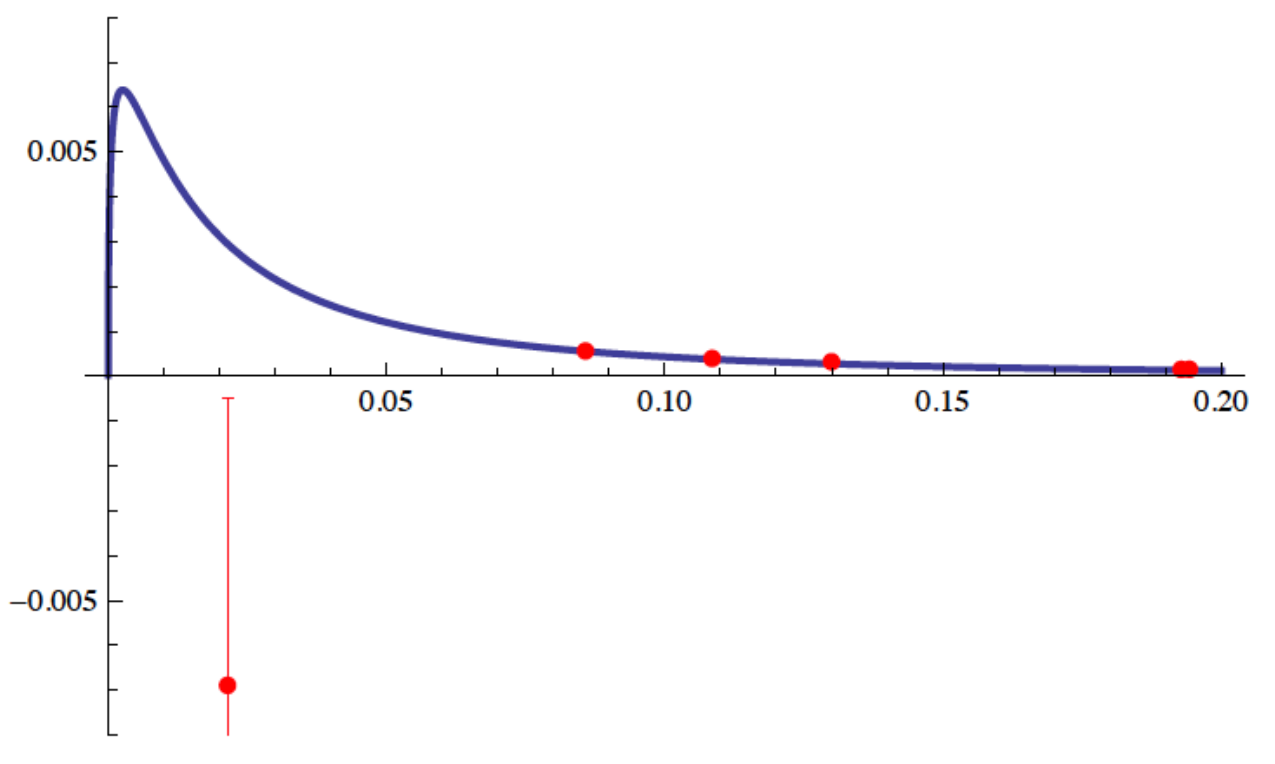}
\hspace{.1cm}
\includegraphics[width=2.9in]{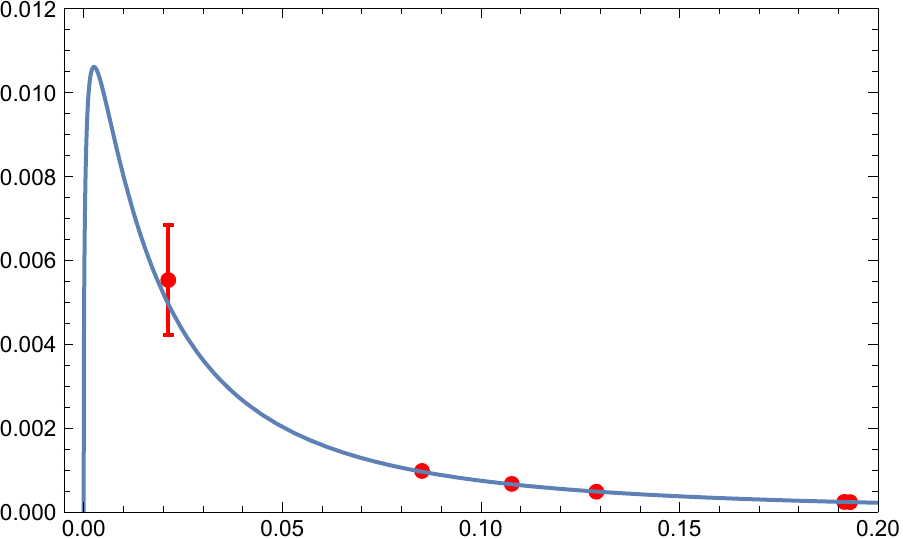}
\label{fig1}
\caption{Typical fits (blue curves) to lattice data (red points)
for the hadronic vacuum polarization. The figure on the left shows data with
statistics typical of 2012 \cite{Pade}, the figure on the right shows data with statistics
typical of 2015 (unpublished).}
\vspace*{2ex}
\end{figure}

The leading-order hadronic contribution to the muon anomalous magnetic moment
$g=2(1+a_\mu)$ is given by the expression
\begin{equation}
\label{amu}
a^{\rm HVP}_\mu=
\left(\frac{\alpha}{\pi}\right)^2\int_0^\infty dQ^2\,f(Q^2)\,[\Pi(Q^2)-\Pi(0)]\ ,
\end{equation}
in which $f(Q^2)$ is a known weight function which depends on the muon mass, and $\Pi(Q^2)$ is defined by
\begin{equation}
\label{vacpol}
\Pi_{\mu\nu}(Q)=\left(\delta_{\mu\nu}Q^2-Q_\mu Q_\nu\right)\Pi(Q^2)\ ,
\end{equation}
the (euclidean) hadronic vacuum polarization \cite{TB2003,ER}.   The integrand
of Eq.~(\ref{amu}) is peaked around $Q^2\sim m_\mu^2/4$, and shown in
Fig.~\ref{fig1}.
The figure on the right (which was obtained with AMA improvement \cite{AMA})
shows that dramatic improvement with statistics was made over the last few
years.   But the figures also show that $a^{\rm HVP}_\mu$ is very sensitive to the
very-low $Q^2$ region.   Thus, in addition to the need for more data with
smaller errors at low $Q^2$, it is also important to understand systematic
effects in detail.   Very small changes in $\Pi(Q^2)$ at low $Q^2$ can have
dramatic effects on the value of $a^{\rm HVP}_\mu$.   Here, we will consider
the impact of finite-volume effects on $a^{\rm HVP}_\mu$, for the case of a
volume $L^3\times T$, with periodic boundary conditions, $L$ the spatial
extent, and $T$ the temporal extent, with $L\ne T$.   

\section{Theoretical considerations}
A first observation is that in finite volume, the Ward--Takahashi identity does
not exclude that $\Pi_{\mu\nu}(0)$ does not vanish, because of the discrete
nature of momenta in a finite volume.\footnote{This observation, as well as
various other observations made below, were also made in Ref.~\cite{BM}.}
Furthermore, it is reasonable to expect that the hadronic vacuum polarization,
with its $\log{Q^2}$ behavior, is more singular for low momenta in a finite
volume than in infinite volume.   This suggests subtracting $\Pi_{\mu\nu}(0)$ 
from the vacuum polarization at non-zero $Q$.   We thus define
\begin{eqnarray}
\label{subPimunu}
\overline{\Pi}_{\mu\nu}(Q)&\equiv& P^T_{\mu\kappa}(Q)\left(\Pi_{\kappa\lambda}(Q)-\Pi_{\kappa\lambda}(0)\right)
P^T_{\lambda\nu}(Q)\ ,\\
P^T_{\mu\nu}(Q)&=&\delta_{\mu\nu}-\frac{Q_\mu Q_\nu}{Q^2}\nonumber\ .
\end{eqnarray}
We chose to project the subtracted vacuum polarization so that it satisfies the
Ward--Takahashi identity, but this turns out not to be essential in what follows.

\begin{figure}[t]
\centering
\includegraphics[width=2.9in]{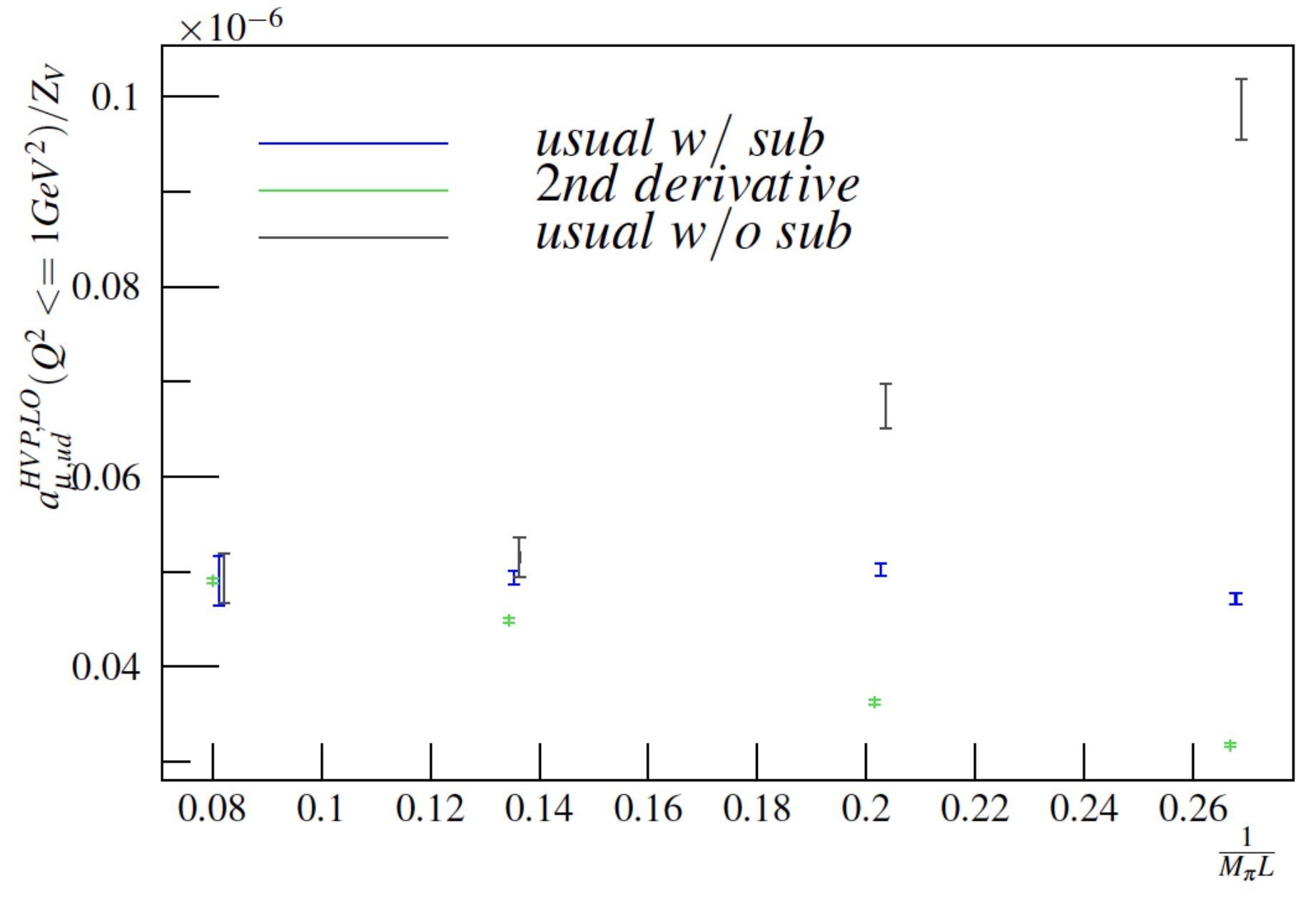}
\label{fig2}
\caption{Dependence on volume of $a^{\rm HVP}_\mu$, from Ref.~\cite{BMW}.
The black points have been obtained without the subtraction of $\Pi_{\mu\nu}(0)$, the blue points with such a subtraction.   These data have $a=0.104$~fm,
$m_\pi=292$~MeV, and $3.7\le m_\pi L\le 12.3$.   For more details, see
Ref.~\cite{BMW}. }
\vspace*{2ex}
\end{figure}

This idea has been considered before.   For instance, in Ref.~\cite{BMW}
a study was made of the effect of this subtraction at different values of the
volume; Fig.~\ref{fig2} shows some of the results.
It is clear that the subtraction reduces finite-volume effects significantly, at least
for the choice of parameters that were used in Fig.~\ref{fig2}.

On the lattice, rotational symmetry is broken by both the lattice itself, and by the
shape of the finite volume.    When rotational symmetry is broken, 
the Ward--Takahashi identity allows for tensor structures other than $Q^2\delta_{\mu\nu}$
and $Q_\mu Q_\nu$ in $\Pi_{\mu\nu}(Q)$, such as for instance
$\delta_{\mu\nu}\sum_\kappa Q^4_\kappa$ and
$Q^3_\mu Q_\nu+Q_\mu Q^3_\nu$.
However, because of dimensions, such terms have to appear with coefficients with
mass dimension $-2$.   That implies that the coefficients have to contain a power
$a^2$, because the other scales in the theory such as $L$ and $m_\pi$ would
have to appear as $L^2$ or $m_\pi^{-2}$, which is clearly impossible. 
Since we are interested in the low-$Q^2$ region, we will assume that, for such
momenta, scaling violations can be ignored.  It follows that $\overline\Pi_{\mu\nu}(Q)$
has to be of the form (\ref{vacpol}), with the proviso that, since the unbroken
group of rotations are just the spatial cubic rotations by 90 degrees, there is 
more than one irreducible representation (irrep) of the cubic group hiding in this
decomposition.   In particular, we may project onto the irreps
\begin{eqnarray}
\label{irreps}
A_1:\qquad &&\sum_i\Pi_{ii}\quad\mbox{and}\quad\Pi_{44}\ ,\\
T_1:\qquad &&\Pi_{4i}=\Pi_{i4}\ ,\nonumber\\
T_2:\qquad &&\Pi_{i\ne j}=\Pi_{j\ne i}\ ,\nonumber\\
E:\qquad &&\Pi_{11}-\sum_i\Pi_{ii}/3\ ,
\Pi_{22}-\sum_i\Pi_{ii}/3\ ,\nonumber
\end{eqnarray}
and extract a scalar function $\Pi(Q^2)$ from each of these.   Since cubic
rotations do not transform the five different irreps into each other, these
scalar functions do not have to be equal.   They only become equal to 
each other in the limit $L,\ T\to\infty$.\footnote{The Ward--Takahashi identity
implies certain relations between the five irreps, for each choice of momentum.}
We will label these five different scalar functions as 
$\Pi_{A_1}$ (from $\sum_i\Pi_{ii}$), $\Pi_{A_1^{44}}$ (from $\Pi_{44}$), $\Pi_{T_1}$, $\Pi_{T_2}$ and $\Pi_E$.

Assuming that finite-volume effects are dominated by pions, one may also
study them in chiral perturbation theory (ChPT).   To leading order in ChPT,
the hadronic vacuum polarization due to pions in a finite periodic volume
$L^3\times T$ is given by
\begin{eqnarray}
\label{vacpolchpt}
&\Pi^{\rm ChPT}_{\mu\nu}(Q)=
\frac{5}{9}e^2\Bigg(4\,\frac{1}{L^3T}\sum_p
\frac{\sin{\left(p+Q/2\right)_\mu}\sin{\left(p+Q/2\right)_\nu}}
{\left(2\sum_\kappa(1-\cos{p_\kappa})+m_\pi^2\right)
\left(2\sum_\kappa(1-\cos{(p+Q)_\kappa})+m_\pi^2\right)}\\
&-2\,\delta_{\mu\nu}\,\frac{1}{L^3T}\sum_p
\left(\frac{\cos{p_\mu}}{\left(2\sum_\kappa(1-\cos{p_\kappa)}+m_\pi^2\right)}
\right)\Biggr)\ ,\nonumber
\end{eqnarray} 
where $e$ is the electric charge of the electron, and the sums are over 
quantized momenta $p_\mu=2\pi n_\mu/L_\mu$, with $n_\mu$ integers,
and $L_1=L_2=L_3=L$ and $L_4=T$.   One can show explicitly from this
expression that $\Pi^{\rm ChPT}_{\mu\nu}(0)$ does not vanish, but rather, that it is
exponentially suppressed with $m_\pi L$.
In our explorations below, we will
compute $\Pi^{\rm ChPT}_{\mu\nu}(Q)$ omitting the factor $5e^2/9$.

It is rather well known that leading-order ChPT does not give a good
description of vector and axial-vector two-point functions already for
very small values of $Q^2$.\footnote{See, for example, Refs.~\cite{AB2007,L10}.}
The intuitive reason for this is that the $\rho$ and $a_1$ resonances
make significant contributions to these two-point functions, whereas leading-order
ChPT only sees the pions (vector resonances
contribute only through low-energy constants at higher order).   However,
here we are only interested in the difference between the vacuum
polarization in finite and infinite volume.   Since for large enough 
values of $m_\pi L$ these differences are exponentially small in the
ratio of the linear volume and hadronic Compton wave lengths, it is
reasonable to assume that these differences are dominated by pions,
and thus well described by leading-order ChPT.   We will investigate this
in what follows.

\section{Comparison between lattice data and theory}
\begin{figure}[t]
\centering
\includegraphics[width=2.9in]{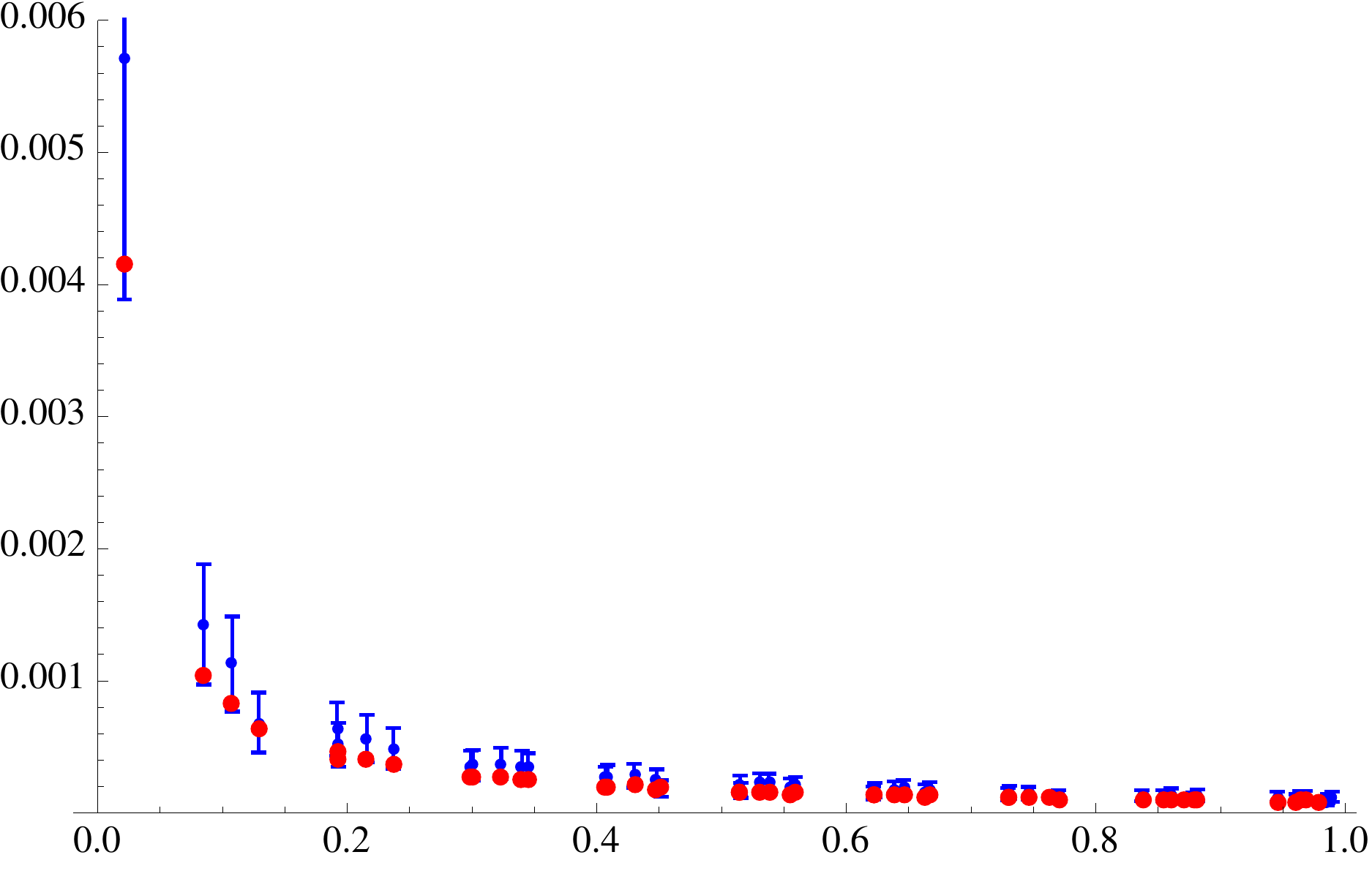}
\hspace{.1cm}
\includegraphics[width=2.9in]{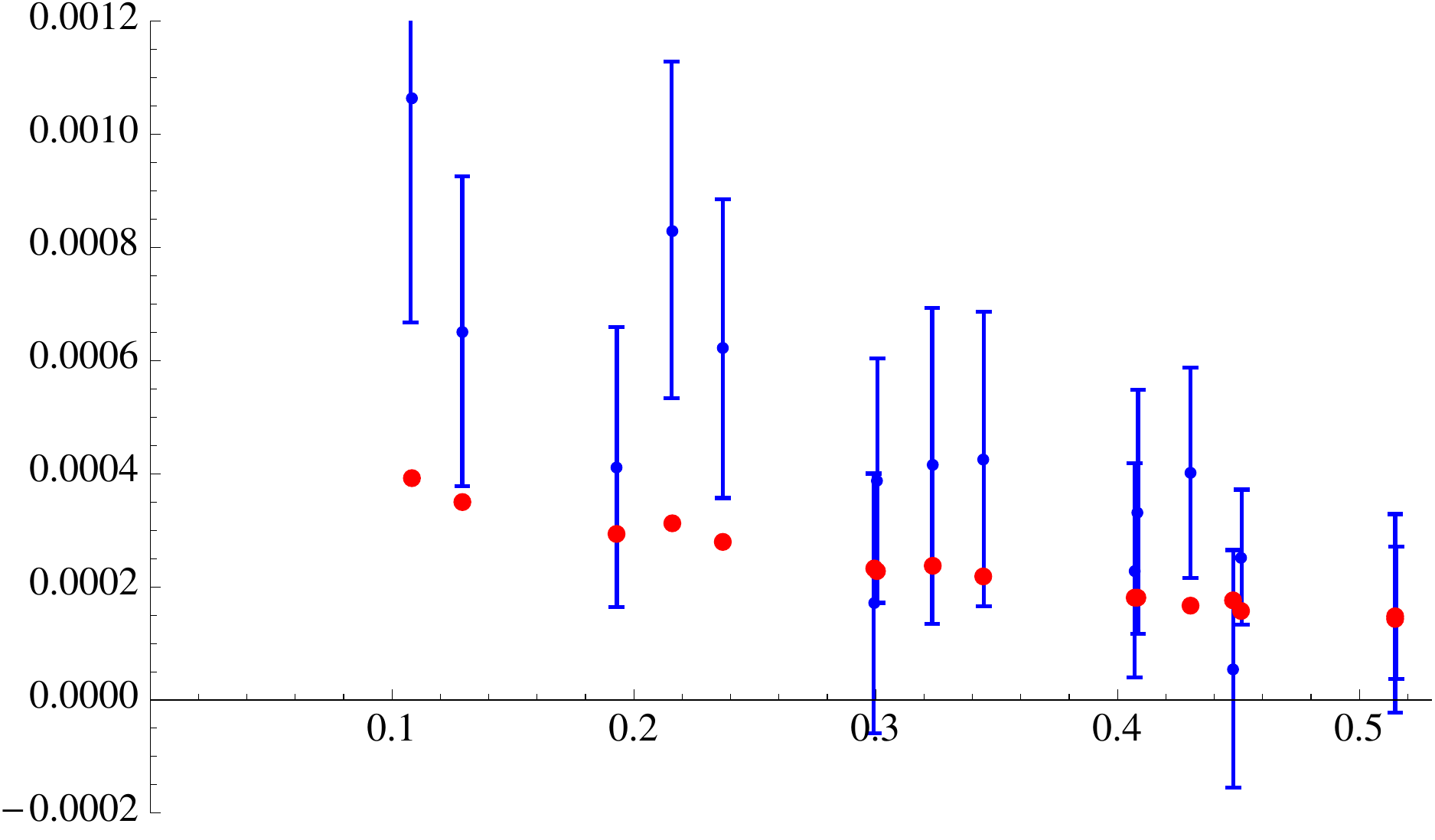}
\label{fig3}
\caption{Left panel:  difference between the subtracted ($\overline\Pi_{A_1}(Q^2)$) 
and unsubtracted ($\Pi_{A_1}(Q^2)$) $A_1$ vacuum polarizations.
Right panel: difference between the subtracted ($\overline\Pi_{A_1}(Q^2)$) 
and unsubtracted ($\Pi_{A_1^{44}}(Q^2)$) $A_1^{44}$ vacuum polarizations.
Red points are computed in ChPT; blue points are lattice data.}
\vspace*{2ex}
\end{figure}
\begin{figure}[t]
\centering
\includegraphics[width=2.9in]{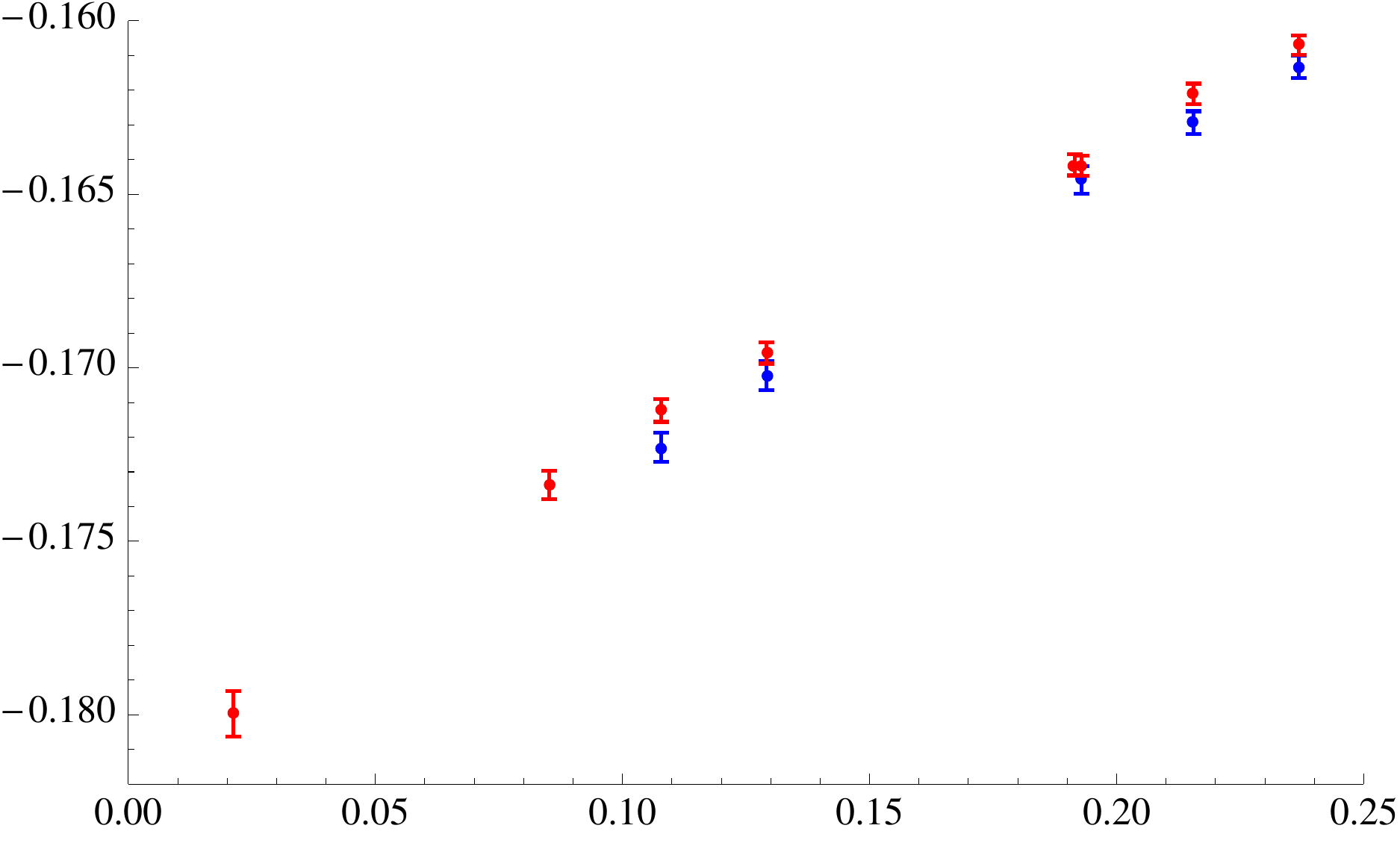}
\caption{Lattice data for $\overline\Pi_{A_1}(Q^2)$
and $\Pi_{A_1^{44}}(Q^2)$. }
\label{fig4}
\vspace*{2ex}
\end{figure}

Figure~\ref{fig3} shows a comparison between ChPT, using Eq.~(\ref{vacpolchpt}), and
lattice data.  The lattice points were computed using the MILC asqtad ensemble
with $1/a=3.34532$~GeV, $m_\pi=220$~MeV, $L=64a$ and $T=144a$, which
implies $m_\pi L=4.2$.   We find that the subtraction of $\Pi_{\mu\nu}(0)$ only
makes a significant difference for $\Pi_{A_1}$ (as one might expect).   The
left panel shows the effect of the subtraction itself, while the right panel shows
the difference between two different irreps.   We see that there is good agreement
between ChPT and lattice data, implying that leading-order ChPT does a 
reasonably good job of describing finite-volume effects.   Similar results are 
obtained for differences between other irreps.   Another important conclusion is that
the lattice data are precise enough to be able to discern finite-volume effects,
thanks to the use of all-mode averaging \cite{AMA}, which was employed to get the data shown.   To illustrate this observation,
we show in Fig.~\ref{fig4} the lattice data points for $\overline\Pi_{A_1}(Q^2)$
and $\Pi_{A_1^{44}}(Q^2)$.

\begin{figure}[t]
\centering
\includegraphics[width=2.9in]{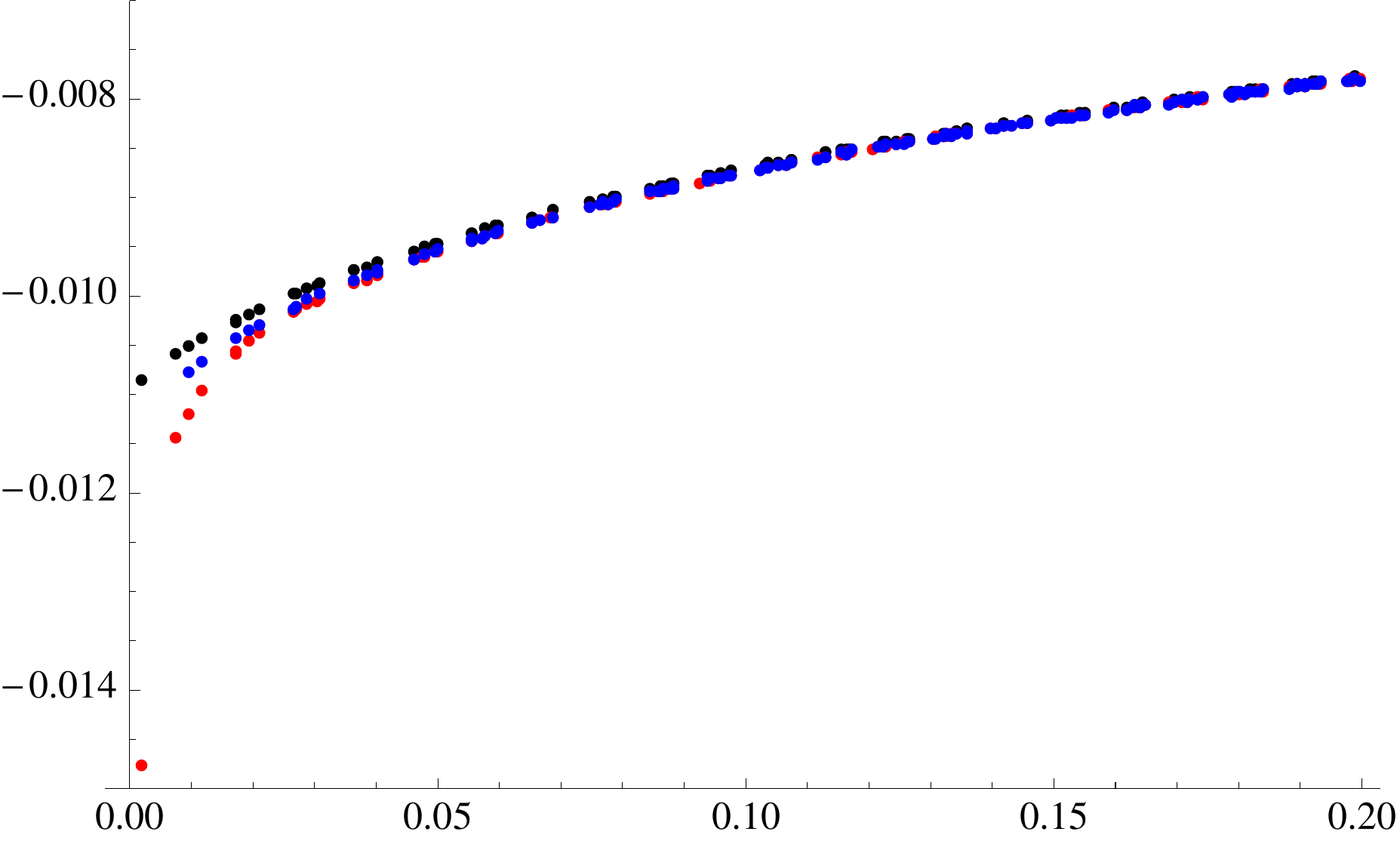}
\hspace{.1cm}
\includegraphics[width=2.9in]{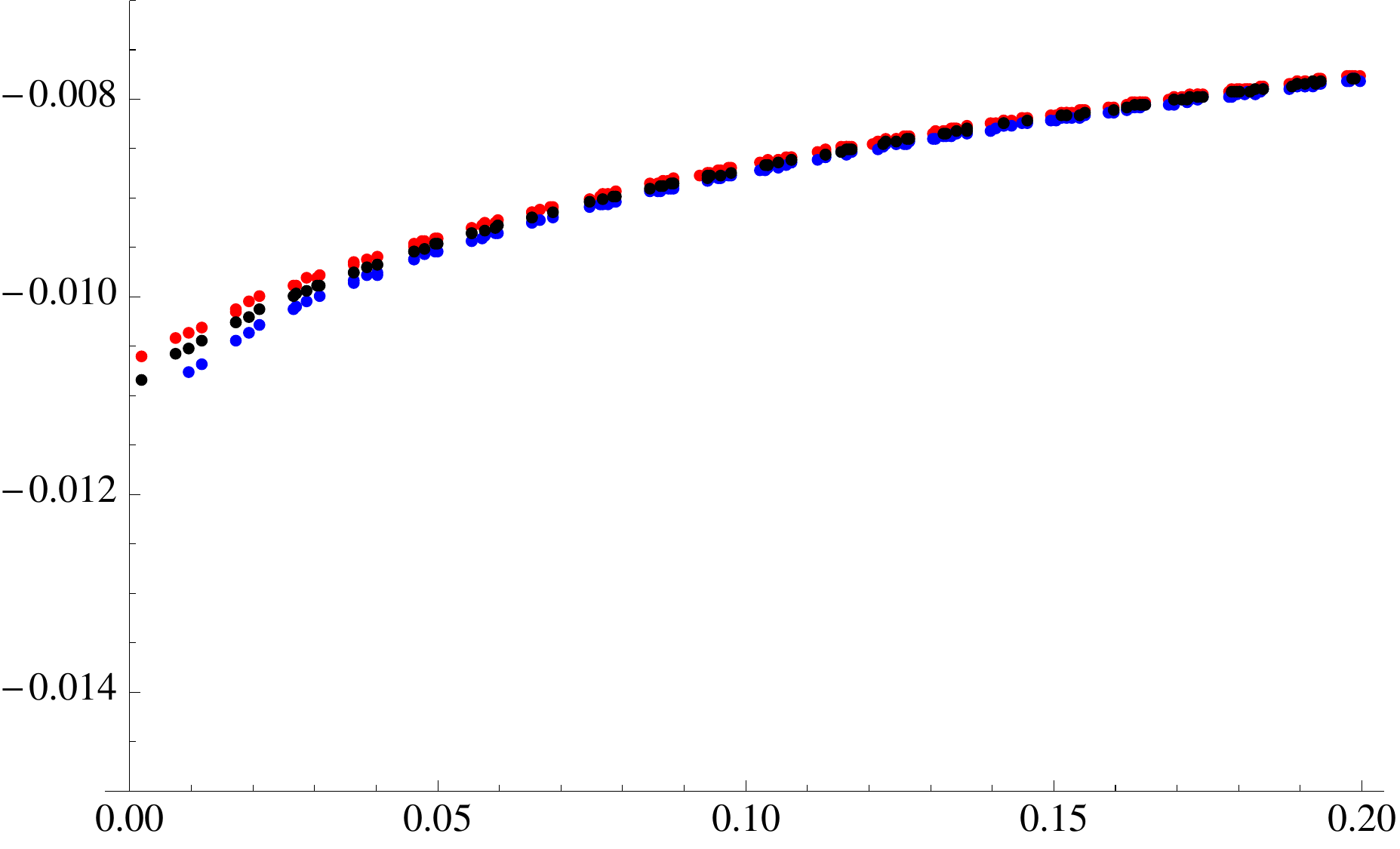}
\caption{Left panel: ChPT computation of (unsubtracted) $\Pi_{A_1}(Q^2)$ 
(red points), (unsubtracted) $\Pi_{A_1^{44}}(Q^2)$ (blue points) and the ``infinite-volume'' result (black points).  Right panel: ChPT computation of (subtracted) $\overline\Pi_{A_1}(Q^2)$ 
(red points), (unsubtracted) $\Pi_{A_1^{44}}(Q^2)$ (blue points) and the ``infinite-volume'' result (black points).}
\label{fig5}
\vspace*{2ex}
\end{figure}

In Fig.~\ref{fig5}, we compare ChPT results for the $A_1$, $A_1^{44}$ and
infinite-volume vacuum polarizations, with the unsubtracted $A_1$ case
on the left, and the subtracted $A_1$ case on the right.   Infinite-volume
points were computed by replacing $L\to 2L$ and $T\to 2T$ in Eq.~(\ref{vacpolchpt}).\footnote{Since the finite-volume effects are exponentially small, the difference
between the black points and the real infinite-volume values is not visible on the
scale of Fig.~\ref{fig5}.}   As mentioned before, there is very little difference
between the subtracted and unsubtracted values for $\Pi_{A_1^{44}}(Q^2)$,
and similar plots for the irreps $T_{1,2}$ and $E$ look very similar.   As one
sees by comparing the left and right panels, the effect of the subtraction
for the $A_1$ irrep is dramatic at the lowest values of $Q^2$, and the subtraction
brings $\Pi_{A_1}(Q^2)$ much closer to the infinite-volume result.   Thus, ChPT
provides a theoretical explanation of this effect.  Moreover, it is interesting that
after the subtraction, the $A_1$ and $A_1^{44}$ vacuum polarizations straddle
the infinite-volume result.   The same thing happens if we replace $A_1^{44}$
by any of the other irreps $T_{1,2}$ and $E$.

\section{Effects on $a_\mu^{\rm HVP}$}
We now consider what the small, but significant finite-volume effects in $\Pi(Q^2)$ imply for
$a_\mu^{\rm HVP}$.   We define
\begin{equation}
\label{amucut}
a^{\rm HVP}_\mu(Q^2_{max})=
\left(\frac{\alpha}{\pi}\right)^2\int_0^{Q^2_{max}}dQ^2\,f(Q^2)\,[\Pi(Q^2)-\Pi(0)]\ ,
\end{equation}
differing from $a^{\rm HVP}_\mu$ because of the choice of upper limit of the
integral.   Below we will choose $Q^2_{max}=1$~GeV$^2$, to diminish the
effect of different
systematic errors at large momenta.   Close to 100\% of the light-quark
contribution to $a^{\rm HVP}_\mu$ comes from the region below 
$Q^2=1$~GeV$^2$ \cite{strategy}.   

We first consider $[0,1]$ Pad\'e fits \cite{Pade} on the interval below 1~GeV$^2$, using
the $A_1$ or $A_1^{44}$ data points only.   We find 
\begin{eqnarray}
\label{pade}
a^{\rm HVP}_{\mu,A_1}(1~\mbox{GeV}^2)&=&8.4(4)\times 10^{-8}\ ,\\
a^{\rm HVP}_{\mu,A_1^{44}}(1~\mbox{GeV}^2)&=&9.2(3)\times 10^{-8}\ .\nonumber
\end{eqnarray}
Using instead a quadratic conformally-mapped polynomial \cite{strategy},
we find
\begin{eqnarray}
\label{conformal}
a^{\rm HVP}_{\mu,A_1}(1~\mbox{GeV}^2)&=&8.4(5)\times 10^{-8}\ ,\\
a^{\rm HVP}_{\mu,A_1^{44}}(1~\mbox{GeV}^2)&=&9.6(4)\times 10^{-8}\ .\nonumber
\end{eqnarray}
While one may argue that these fit functions are not adequate to reach the
desired sub-percent level accuracy, the point here is that the Pad\'e and
conformally-mapped polynomial fits give results that are consistent within
errors (of order 4\%).   The difference between the $A_1$ and $A_1^{44}$
fits, however, is about 9--13\% larger than the errors shown in Eqs.~(\ref{pade},\ref{conformal}), and consistent between the two types of fits.   Since the
only difference is the irrep onto which we projected the lattice data, we conclude that
the 9--13\% difference is a finite-volume effect.   A phenomenological analysis
of finite-volume effects found values consistent with our estimate \cite{Mainz}.

\section{Conclusion}
It is difficult to perform a lattice computation of $a^{\rm HVP}_\mu$
with sub-percent level accuracy because of the nature of the integrand
in the integral defining this quantity, as demonstrated in Fig.~\ref{fig1}.
The integrand is strongly peaked at $Q^2\approx m_\mu^2/4$, values
of the momenta which are hard to reach on the lattice.   
Consequently, even small variations of the vacuum polarization itself, due to systematic
errors, get magnified to be large variations on $a^{\rm HVP}_\mu$.
In this talk, we demonstrated that this is also true for the systematic
error due to the use of a finite volume.  

Together with the conclusions reached in Refs.~\cite{Pade,strategy,taumodel}, the picture that emerges is that for good control of the systematic errors, one needs a series of
model-independent fit functions to the lattice data for $\Pi(Q^2)$, 
approximately physical pion masses, and control over finite-volume
effects.

Finally, we note that finite-volume effects are equally likely to affect
other methods to construct smooth interpolations of the lattice data
for $\Pi(Q^2)$, such as the moment method proposed in Ref.~\cite{HPQCD}.
In particular, in a finite volume, the $t^2$ moment of the vector current
correlator is not equal to $\Pi(0)$, but instead to a linear
combination of values at non-zero momenta \cite{BI}:
\begin{equation}
\label{moment}
\Pi(0)\to\sum_{n\ne 0}4(-1)^n\Pi\left(\frac{2\pi n}{T}\right)\ .
\end{equation}
We do not know of any argument that this linear combination of values 
of $\Pi(Q^2)$ is less sensitive to finite-volume effects than any of these
values alone.

\vspace{2ex}
\noindent{\bf Acknowledgments}
We thank Taku Izubuchi and Kim Maltman for useful discussions.   
TB, PC, MG and CT were supported in part by the US Dept. of
Energy; SP  by CICYT-FEDER-FPA2014-55613-P,
2014~SGR~1450, the Spanish Consolider-Ingenio 2010 Program
CPAN (CSD2007-00042).


\vspace{-1ex}
\begin{thebibliography}{99}
\bibitem{TB2003}
  T.~Blum,
  %``Lattice calculation of the lowest order hadronic contribution to the muon anomalous magnetic moment,''
  Phys.\ Rev.\ Lett.\  {\bf 91}, 052001 (2003)
  [hep-lat/0212018].
  %%CITATION = HEP-LAT/0212018;%%
  
\bibitem{ER}
  B.~E.~Lautrup, A.~Peterman and E.~de Rafael,
  %``Recent developments in the comparison between theory and experiments in quantum electrodynamics,''
  Phys.\ Rept.\  {\bf 3}, 193 (1972).
  %%CITATION = PRPLC,3,193;%%

\bibitem{Pade} 
  C.~Aubin, T.~Blum, M.~Golterman and S.~Peris,
  %``Model-independent parametrization of the hadronic vacuum polarization and g-2 for the muon on the lattice,''
  Phys.\ Rev.\ D {\bf 86}, 054509 (2012)
  [arXiv:1205.3695 [hep-lat]].
  %%CITATION = ARXIV:1205.3695;%%

\bibitem{AMA} 
  T.~Blum, T.~Izubuchi and E.~Shintani,
  %``New class of variance-reduction techniques using lattice symmetries,''
  Phys.\ Rev.\ D {\bf 88}, no. 9, 094503 (2013)
  [arXiv:1208.4349 [hep-lat]].
  %%CITATION = ARXIV:1208.4349;%%
  
\bibitem{BM}
  D.~Bernecker and H.~B.~Meyer,
  %``Vector Correlators in Lattice QCD: Methods and applications,''
  Eur.\ Phys.\ J.\ A {\bf 47}, 148 (2011)
  [arXiv:1107.4388 [hep-lat]].
  %%CITATION = ARXIV:1107.4388;%%
  
\bibitem{BMW}
  R.~Malak {\it et al.} [Budapest-Marseille-Wuppertal Collaboration],
  %``Finite-volume corrections to the leading-order hadronic contribution to $g_\mu-2$,''
  PoS LATTICE {\bf 2014}, 161 (2015)
  [arXiv:1502.02172 [hep-lat]].
  %%CITATION = ARXIV:1502.02172;%%
  
\bibitem{AB2007}
  C.~Aubin and T.~Blum,
  %``Calculating the hadronic vacuum polarization and leading hadronic contribution to the muon anomalous magnetic moment with improved staggered quarks,''
  Phys.\ Rev.\ D {\bf 75}, 114502 (2007)
  [hep-lat/0608011].
  %%CITATION = HEP-LAT/0608011;%%
  
\bibitem{L10}
   D.~Boito, M.~Golterman, M.~Jamin, K.~Maltman and S.~Peris,
  %``Low-energy constants and condensates from the $\tau$ hadronic spectral functions,''
  Phys.\ Rev.\ D {\bf 87}, no. 9, 094008 (2013)
  [arXiv:1212.4471 [hep-ph]].
  %%CITATION = ARXIV:1212.4471;%%
  
\bibitem{strategy}
  M.~Golterman, K.~Maltman and S.~Peris,
  %``New strategy for the lattice evaluation of the leading order hadronic contribution to $(g?2)\mu$,''
  Phys.\ Rev.\ D {\bf 90}, no. 7, 074508 (2014)
  [arXiv:1405.2389 [hep-lat]].
  %%CITATION = ARXIV:1405.2389;%%

\bibitem{Mainz}
  A.~Francis, B.~Jaeger, H.~B.~Meyer and H.~Wittig,
  %``A new representation of the Adler function for lattice QCD,''
  Phys.\ Rev.\ D {\bf 88}, 054502 (2013)
  [arXiv:1306.2532 [hep-lat]].
  %%CITATION = ARXIV:1306.2532;%%
  
\bibitem{taumodel}
  M.~Golterman, K.~Maltman and S.~Peris,
  %``Tests of hadronic vacuum polarization fits for the muon anomalous magnetic moment,''
  Phys.\ Rev.\ D {\bf 88}, no. 11, 114508 (2013)
  [arXiv:1309.2153 [hep-lat]].
  %%CITATION = ARXIV:1309.2153;%%
  
\bibitem{HPQCD}
  B.~Chakraborty {\it et al.} [HPQCD Collaboration],
  %``Strange and charm quark contributions to the anomalous magnetic moment of the muon,''
  Phys.\ Rev.\ D {\bf 89}, no. 11, 114501 (2014)
  [arXiv:1403.1778 [hep-lat]].
  %%CITATION = ARXIV:1403.1778;%%
  
\bibitem{BI}
  T.~Blum and T.~Izubuchi, private notes (2015).

\end{thebibliography}
\end{document}